# Fully Digital: Policy and Process Implications for the AAS

Chris Biemesderfer, American Astronomical Society

Over the past two decades, every scholarly publisher has migrated at least the mechanical aspects of their journal publishing so that they utilize digital means. The academy was comfortable with that for a while, but publishers are under increasing pressure to adapt further. At the American Astronomical Society (AAS), we think that means bringing our publishing program to the point of being fully digital, by establishing procedures and policies that regard the digital objects of publication primarily. We have always thought about our electronic journals as databases of digital articles, from which we can publish and syndicate articles one at a time, and we must now put flesh on those bones by developing practices that are consistent with the realities of article at a time publication online. As a learned society that holds the long-term rights to the literature, we have actively taken responsibility for the preservation of the digital assets that constitute our journals, and in so doing we have not forsaken the legacy pre-digital assets. All of us who serve as the long-term stewards of scholarship must begin to evolve into fully digital publishers.

## Introduction

Scholarly communication has evolved in interesting ways for millennia. From Archimedes writing letters on scrolls to his correspondents, to Neil deGrasse Tyson tweeting to his followers, scholars have discussed not only the issues of their disciplines and the insights of their research, they have (at least recently) discussed how other scholars communicate as well (Odlyzko 2002, Heuer et al. 2008, Renear and Palmer 2009). However, in this paper I won't talk about blogs, wikis, txts, tweets, youtubes, crowd-sourced GPS-enabled climate-sensitive thoroughly-modern mobile droidbots – and all that jazz. I intend to confine my remarks to *formal* communications: information that is published in the things we recognize today as the peer-reviewed journals. It is worth stating plainly that at the AAS, we are primarily concerned with communications that benefit professional astronomers. For the purpose of this paper, that means I won't worry about how we might address audiences other than researchers.

I'm also going to assume that the notion of formal communication as we know it is basically a good thing, because it comprises other good things: peer review is a good thing, consistency is a good thing, permanence (durability, longevity) is a good thing. I acknowledge that there are voices crying out against some of these things – "peer review must die", e.g. – and certainly there is nothing wrong with questioning fundamental tenets. But let me make an observation.

Archimedes wrote to his colleagues in a form that is structurally very much like communications between scholars today. We can read modern translations of Archimedes, and they seem familiar to us; the same is true for works of Euclid or Eratosthenes or Galileo. The information has certainly been durable (although in the case of Archimedes, not without some close calls; Netz and Noel, 2007). What is remarkable, however, is how consistent the form is, whether written in parchment scrolls, bound codices, printed books, or LaTeX files. In spite of hyperventilated claims that "digital changes everything", formal communication has survived format changes intact for thousands of years. My point is not to belittle Web 2.0 style communications; my point is not that formal communications are better somehow; my point is that formal communication is not going away.

As a specific instance of formal communication, scholarly journals have a number of merits for the organization of these communications. They provide an initial selection of articles through different editors with different aims and scopes; they offer a range of editorial temperaments; they provide peer review; and they are a platform for the methodical organization, normalization, and preservation of information. In addition, as a publisher of scholarly journals, the AAS offers scholars a sensible approach to intellectual property, and the research community a rational and sustainable business model. Every one of these things – every single one – is technology-neutral. We've done them all along, we're still doing them, and at the American Astronomical Society, we think we should keep doing them.

## Publishing initiatives at the AAS

The AAS has actively pursued innovation and modernization in its publications. Indeed, the venerable *Astrophysical Journal* (ApJ) itself is sometimes described as having been born in order to communicate the new scientific insights that were forming in the wake of the invention of spectroscopy (Osterbrock 1995). When the ApJ needed to publish more data tables, a Supplement series was created in 1954. When speedier publication of new results was indicated, the Letters were published on a faster schedule (Chandrasekhar 1967). When technology drove us beyond the printed page, the ApJ added videos in 1992, and the AAS began to distribute digital data on CD-ROM in 1993.

The AAS began to take its publications in their entirety into the digital regime with the abstracts for the Society's 180$^{th}$ meeting in Columbus in 1992; by the 182$^{nd}$ meeting in Berkeley in 1993, the abstracts were published on the World Wide Web. The Society had embarked on collaborations, also in 1992, with its publisher at the time, The University of Chicago Press, and with a project called STELAR (Study of Electronic Literature for Astronomical Research) which was associated with the National Space Science Data Center at Goddard Space Flight Center (Warnock 1993). After a period of morphological instability and accretion, STELAR gave way to ADS, the Astrophysics Data System (Kurtz et al. 2000), which had been developing independently in another part of the universe (Cambridge, Massachusetts). The large research journals followed into the digital realm through the mid-1990s (Boyce 1998), and after nearly twenty years, we are preparing once again to update the *Bulletin of the AAS* (which has served as the print home for the meeting abstracts).

The present-day journal publishing initiatives at the AAS have three main strategic drivers: to provide more underlying numerical materials in the journals, to manage the evolution of print, and to adjust the business model accordingly. Since 1990, the Society and its publishing partners have effected the "digitalization" of all the major facets of the publication process: manuscript preparation and submission by authors, peer review, production and delivery, and preservation.

In the near term, we anticipate that enterprise-scale printing will be phased out in the next 2-3 years, as the library subscribers to the journals stop acquiring the print products. We are going to be looking to web-to-print solutions so that customized print products can be specified by the customers themselves, thus allowing the AAS to focus on the larger issues of professional scholarly communication. In the meantime, we are thinking about a business model for the Society that offers only online subscriptions, and we are preparing to charge authors in 2011 based on the quantities of digital material that are submitted, rather than based on the number of typeset pages of the authors' text.

The AAS' interest in providing data sets explicitly dates back to at least 1954, when the ApJ Supplement was born. In the age of computer networks, the research environment has been significantly enriched by direct live connections among resources. We can no longer limit our attention to data "in the journal" or attached explicitly to articles. Now, journal articles can refer to raw data held in archives and data centers, either at the author's initiative or through the addition of query tools in the online article.

**The view toward the future**

There are so many exciting things we might do with our journals to enrich formal communication among scholars. We are all interested in connecting our journals' consumers with the full range of resources that support (and may extend) the research they are reading about. We have made some good steps already, but what we have accomplished with our journals to date has been done with a relatively small number of partners, mostly through bilateral collaborations. That's good, but it isn't sustainable. If we want to create a healthy and stable and scalable formal scholarly communication enterprise in the future, we have to put systems in place that utilize standards, because the alternative – having to manage NxN system interfaces through bilateral agreements – is fiscally intractable.

Efforts to deliver more machine-readable data in the journals will resonate with energies now being applied to next-generation web technologies that are intended to facilitate machine interactions: the Semantic Web, or the computable web. We don't have to build the applications for the journals, and at the AAS we don't plan to; we just have to make our content accessible via transparent interfaces so the data can be harvested by users for their preferred applications and services. Even doing so, there are crucial questions regarding data that we obtain with journal articles. 1) How should we obtain it, and what are agreeable formats to get from authors? 2) How can we improve the online presentation, especially for complex data objects? What does a "good" user interface for databases look like? 3) For preservation purposes, is any policy more proactive than byte preservation feasible? Where should this kind of data reside for curation? Should it be the same place as for delivery, or might distinct repositories be preferable?

For some purposes, the underlying data is actually the article text. We have seen an increase in interest in mining the text of the journals themselves as a data set, and we anticipate more of this will happen as people and their software agents become more facile with large corpuses of text. We could argue that many of the rendering enhancements we envision for the online journals qualify as attempts to deliver essential information elements as if they were data. For instance, presenting the math via MathJax can be seen as making the math itself more computable.

Publishers have all done respectable enough jobs of "digitalizing" the major elements of the formal communication process. There is a host of fairly unspectacular tasks left to do, the details in which devils reside. Some of tasks are obviously complicated, such as making all those live connections to external data resources. Some of them seem straightforward, but aren't. For instance, switching a paginated journal so that it utilizes article numbering sounds easy, but it is fiendishly complicated because it affects bibliographic coordinates and every system that uses them (Chaix 2010).

Much of the work we have to do going forward will be difficult and time-consuming, mostly because it involves so many other parties, parties that we are obliged to work with. It's the price of interconnectedness. The interconnectedness of the present and future holds promise as well as many challenges. In principle, everything interacts; that's the promise. In practice, everything *has* to interact; that's the challenge, to be able to find partners and techniques that permit the interconnections to arise in ways so that programs are sustainable.

But what concerns me most is that our view toward the future has to be much larger, much more inclusive and comprehensive, than it is today. We face significant challenges, and after a few decades of evolution of the network, we have to confront a host of external pressures that are *not* technological, and quite a few of them at the same time.

Some of the external pressures are not new. Publishers of the scholarly literature are concerned with broad dissemination, and we routinely involve secondary providers to aid in discovery and delivery. We have to engage the "new secondaries", which in astronomy means arXiv and ADS for the most part, but in an inclusive and comprehensive world also has to include individual institutional repositories too numerous to count. There are government mandates; open access comes to mind first, but mandatory deposit of digital content is a non-trivial consideration for the national libraries that require it. The scholarly literature is long-lived, and we will always be concerned with conservation and preservation.

In the future, metrics like the impact factor *might* persist for journals, but it seems reasonable that these may decline in favor of metrics that focus on individuals and teams and research organizations. The community is trying hard to bridge that gap, although not with much success I dare say. Nevertheless, I expect these problems will be solved someday, and we will be able to judge the quality of the contributions of organizations or departments or campuses, and of different research teams and coalitions, and maybe even of individual scientists. At that point, journals will no longer proxy for such entities, and the impact factor et al. will not be so interesting. At the same time, journals will instead be favored for things like speed of publication, equanimity of peer review, and the transparency of the journal's databases to processes that *compute* individual metrics and scores. In the meantime, we will continue to confront statistics that are poorly conceived or unreliable (Adler 2008, Arendt 2010).

Some of the external pressures are unrelated to scholarly publishing. Information overload will get worse before it never gets better. And this evolution in formal communication is happening in a global technological environment that appears to most people to be quite easily switched on, thanks to the effectiveness of advertising and promotion in our modern digital world. As a result, our customers keep asking questions like "How hard can it be?", and our genteel scholarly dialog is drowned out by crowd-sourced cacophony. We will be defending the value propositions of formal communication for a long time.

Let me come back to the idea that our view toward the future has to be more inclusive and comprehensive. In many ways, the world has gotten smaller as globalization has come about. Communication of huge quantities of information is virtually instantaneous thanks to countless kilometers of optical fiber and phalanxes of orbiting satellites. It is possible to travel almost anywhere on the surface of the Earth, quite quickly and inexpensively. These changes have happened fairly rapidly, and in a short span – much less than a lifetime – we have enjoyed a dramatic expansion of our viewpoints. (That larger number of viewpoints and opportunities, by the way, also results in many more sources of information to manage, and more partners to interact with. So in terms of our potential collaborations, the world has also gotten quite a bit bigger.)

However, enjoying expanded viewpoints is not the same as a broadened perspective. Our policies are catching up, but most of our assumptions about the cultural norms that govern the behavior of scholars are rooted in a homogeneous past. The formal communication that has been discussed in this essay is all Western, or at least not any more Eastern than Constantinople. The

Biemesderfer, Fully Digital                                                                                                                6 big small world in which we now communicate challenges us, and thinking only about the West isn't thinking big enough. At the AAS, we are cognizant of the need to adjust our scope and our sensibilities so that we accommodate research and economic engines in Asia, although I think it is fair to say that we do not yet know what all the ramifications of these adjustments will be. However, I do believe that it is a critical part of becoming a fully digital publisher that we are capable of embracing scholarly practices the world around.

We have certainly embraced the technological innovations of the past several decades, the academy has started to adjust attitudes and policies about scholarly communication, and we are beginning to recognize how crucial a broader perspective will be. As fully digital publishers, aided by new technologies and new policies and a broad perspective, we gain the attitudes and the flexibility to meet the present and future challenges of scholarly communication. All of us who serve as the long-term stewards of scholarship need to evolve into fully digital publishers.

I'm grateful for careful readings of drafts by Peter Boyce and Rebecca Jensen; this paper benefits from their suggestions. The AAS has benefited from suggestions and help from all quarters for many years. Recently, I've been able to have stimulating discussions with the AAS editorial and development teams, and in the Executive Office with our Working Group on Communications. Over the years, I've enjoyed fruitful conversations with many innovative colleagues in publishing. And we all have the pleasure of working with a research community full of creativity and new ideas.